\documentclass[10pt,journal,compsoc]{IEEEtran}
\usepackage[nocompress]{cite}
\usepackage{mathtools}
\usepackage[utf8]{inputenc} % allow utf-8 input
\usepackage[T1]{fontenc}    % use 8-bit T1 fonts
\usepackage{hyperref}       % hyperlinks
\usepackage{url}            % simple URL typesetting
\usepackage{booktabs}       % professional-quality tables
\usepackage{amsfonts}       % blackboard math symbols
\usepackage{nicefrac}       % compact symbols for 1/2, etc.
\usepackage{microtype}      % microtypography
\usepackage[caption=false]{subfig}
\usepackage{subfloat}
\usepackage{tabularx}
\usepackage{amsmath}
\usepackage{graphicx} % takes care of graphic including machinery
\usepackage{enumitem}
\usepackage{algorithm, algpseudocode}
\DeclareMathOperator*{\argmax}{arg\,max}

\makeatletter
\newcommand{\algmargin}{\the\ALG@thistlm}
\makeatother
\algnewcommand{\parState}[1]{\State%
	\parbox[t]{\dimexpr\linewidth-\algmargin}{\strut\hangindent=\algorithmicindent \hangafter=1 #1\strut}}

\usepackage{atbegshi}% http://ctan.org/pkg/atbegshi
\ifCLASSINFOpdf
\else
\fi
\begin{document}


\title{Optimizing Online Matching for Ride-Sourcing Services with Multi-Agent Deep Reinforcement Learning}


\author{Jintao~Ke,
        Feng~Xiao,
        Hai~Yang,
        and~Jieping~Ye,~\IEEEmembership{Senior Member,~IEEE}% <-this % stops a space
\IEEEcompsocitemizethanks{
\IEEEcompsocthanksitem J. Ke and H. Yang are with the Department
of Civil and Environmental Engineering, Hong Kong University of Science and Technology, Hong Kong, China. \protect\\
E-mail: jke@connect.ust.hk, cehyang@ust.hk. 
\IEEEcompsocthanksitem F. Xiao (corresponding author) is with School of Business Administration, Southwestern University of Finance and Economics, Chengdu, China.\protect\\
E-mail: evan.fxiao@gmail.com
\IEEEcompsocthanksitem J. Ye . is with AI Labs, Didi Chuxing, Beijing, China\protect\\
E-mail: yejieping@didiglobal.com}
% <-this % stops a space
\thanks{Manuscript submitted March 10, 2019}}


\markboth{}%
{Ke \MakeLowercase{\textit{et al.}}: Optimizing online matching for ride-sourcing services with multi-agent deep reinforcement learning}





\IEEEtitleabstractindextext{%
\begin{abstract}
Ride-sourcing services are now reshaping the way people travel by effectively connecting drivers and passengers through mobile internets. Online matching between idle drivers and waiting passengers is one of the most key components in a ride-sourcing system. The average pickup distance or time is an important measurement of system efficiency since it affects both passengers’ waiting time and drivers’ utilization rate. It is naturally expected that a more effective bipartite matching (with smaller average pickup time) can be implemented if the platform accumulates more idle drivers and waiting passengers in the matching pool. A specific passenger request can also benefit from a delayed matching since he/she may be matched with closer idle drivers after waiting for a few seconds. Motivated by the potential benefits of delayed matching, this paper establishes a two-stage framework which incorporates a combinatorial optimization and multi-agent deep reinforcement learning methods. The multi-agent reinforcement learning methods are used to dynamically determine the delayed time for each passenger request (or the time at which each request enters the matching pool), while the combinatorial optimization conducts an optimal bipartite matching between idle drivers and waiting passengers in the matching pool. Two reinforcement learning methods, spatio-temporal multi-agent deep Q learning (ST-M-DQN) and spatio-temporal multi-agent actor-critic (ST-M-A2C) are developed. Through extensive empirical experiments with a well-designed simulator, we show that the proposed framework is able to remarkably improve system performances.
\end{abstract}

}

\maketitle

% is not selected <OR> if conference mode is selected - because compsoc
% conference papers position the abstract like regular (non-compsoc)
\IEEEdisplaynontitleabstractindextext


% \fi
\IEEEpeerreviewmaketitle

\section{Introduction}
\label{sec:introduction}



\IEEEPARstart{W}{ith} the rapid development of mobile internet based technologies and the popularity of smart phones, ride-sourcing services, such as Uber, Lyft, and DiDi, has been reshaping the way we travel. In contrast to traditional taxi markets that rely on the physical meeting between drivers and passengers, the ride-sourcing platforms implement an on-line matching process that can match unserved passengers and idle drivers more efficiently. Besides, the tremendous amount of information collected through the APPs, such as the real-time locations, trajectories and travel patterns of passengers and drivers, can further enhance the on-line matching procedure by reducing the search frictions between passengers and drivers. There are two main modes of matching process \cite{yang2018universal,xu2018large}. One is the broadcast mode, in which the ride-sourcing platforms collect requests from passengers and broadcasts them to nearby idle drivers, each of whom opts for one of the requests by considering his/her individual benefits (such as trip fare) and costs (such as the distance to pick-up the passenger). Another is the dispatch mode, in which the ride-sourcing platforms collect the information of idle drivers and requests of passengers on the fly and implement a centralized algorithm to match these drivers and passengers pair by pair. Among these two modes, the dispatch mode is thought to be more efficient and thus widely adopted in various ride-sourcing platforms. For example, DiDi’s statistics showed that a significant improvement on system efficiency - the request completion rate being improved by more than 10\% - was observed when DiDi switched from broadcast mode to dispatch mode \cite{xu2018large}.

\begin{figure}
  \centering
  \includegraphics[width=1\columnwidth]{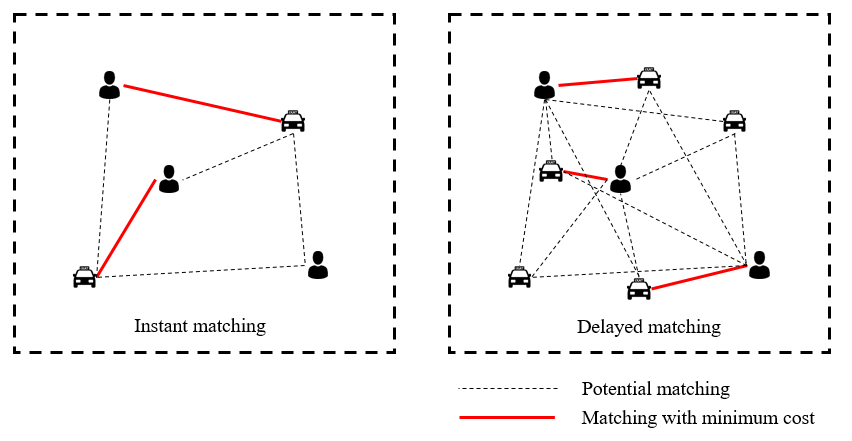}
  \caption{Potential benefits of delayed matching}
  \label{Fig 1:Delayed}
\end{figure}

In dispatch mode, during each short match time interval (such as one or two seconds), the ride-sourcing platforms collect all idle drivers and unserved passenger requests, and then execute matching between drivers and passengers with combinatorial optimization approaches. These approaches aim to improve the overall system performance by increasing the matching rate (the number of matching driver-passenger pairs per unit time), decreasing the passengers’ waiting time (from announcing a request to being matched by the on-line platform), and reducing the passengers’ pick-up time (from being matched to dropping on the corresponding vehicle). Of particular importance here is the trade-off among pick-up time, waiting time, and success rate of matching. When the platforms wait for a few match time intervals to accumulate sufficient pairs of idle drivers and unserved passengers, instead of immediately matching them, a better matching are expected, and as a result, the average pick-up time will be reduced. As shown in Fig. ~\ref{Fig 1:Delayed}, if a delayed matching is used, more drivers and passengers will occur in the matching pool, and thus a more efficient matching between drivers and passengers with shorter average pickup time can be implemented. As for a single passenger, he/she may be matched with a much closer idle driver (much lower pick-up distance) if his/her request is delayed for a few match time intervals by the platform. However, holding passengers and drivers for too many match time intervals brings a long passengers’ waiting time, under which some impatient passengers may cancel their requests. With no doubt, determining the delayed time (or the number of match time intervals to wait) for each passenger request is essential and has potentials to well balance these trade-offs. In reality, with the help of high-efficient computing resources, it is applicable for the ride-sourcing platforms to dynamically determine the delayed time of each request in response to the real-time supply-demand conditions.

Nevertheless, in a highly stochastic and dynamic system, in which passengers’ requests and idle drivers pop up and disappear at any time, to simultaneously determine an optimal delayed time for each passenger request is non-trivial. Although rich historical and real-time information is available, the decisions on the delayed time of each request will impact the future supply-demand conditions (such as the number of unserved requests left from the previous match time interval), which in turn affects the decisions in the following match time intervals. The dynamic interactions between the decisions and the environment (i.e. the supply-demand conditions) cannot be directly characterized by the widely used supervised learning methods. One solution to this problem is the reinforcement learning (RL) \cite{sutton1998introduction} that learns a policy of sequential decision makings by interacting with a complex environment. Recent years have witnessed tremendous success of RL in many transportation problems, such as taxi fleet management \cite{lin2018efficient} \cite{oda2018movi}, signal control \cite{wei2018intellilight}, and adaptive routing problems \cite{mao2018reinforcement}. However, with the traditional RL approaches that use a centralized agent, the action space that includes the decisions for the delayed time of all passengers will become extremely large. 

To tackle these issues, we model the dynamics of the dispatch system as a multi-agent Markov decision process (MMDP), in which each passenger request is viewed as an agent. During each match time interval, each agent decides to be delayed or not to be delayed towards the next time interval. If it is delayed, it will not join the matching pool during this match time interval (simultaneously, one match time interval is added to its delayed time); otherwise, it will join the matching pool, then a bipartite matching between unserved requests and idle drivers in the matching pool will be executed with a combinatorial optimization algorithm. Clearly, under this setting, the action space of each agent reduces to a binary decision (delayed or not) in each match time interval, which makes the originally intractable problem available to be solved. 

The problem setting of this paper is different from the recent related studies in this domain, such as \cite{xu2018large}, \cite{lin2018efficient}, \cite{jindal2018optimizing}, \cite{li2019efficient}, \cite{wang2018deep}, in the following aspects. These studies majorly focused on employing reinforcement learning methods to control the dispatching from orders to drivers and drivers' movements. However, none of these studies considered the potential benefits of delayed matching or investigated the dynamic control of matching time intervals. Moreover, most of these studies treated drivers as agents and tried to optimize the long-term revenue of drivers, whereas our paper focus more on the passenger side, and try to minimize passengers' pickup time through a dynamic control of delayed matching. 

The main contributions of this paper are listed below: 
\begin{itemize}
 \item We propose a two-stage framework that first determines the delayed time of each passenger request through multi-agent reinforcement learning methods and then executes optimal matching between idle drivers and unserved passengers in the matching pool with a combinatorial optimization model.
    \item We formulate the delayed matching problem as a MMDP by a proper design of agent, state, action, and reward. Two types of rewards, global reward and individual reward, are used and then tested. 
    \item We propose two learning methods, named spatio-temporal multi-agent deep Q-learning (ST-M-DQN) and spatio-temporal multi-agent actor-critic (ST-M-A2C), which observe the spatio-temporal patterns of supply and demand and find optimal policies that make a sequence of delayed-or-not decisions for each agent.
    \item We build a simulator that characterizes passengers’ and drivers’ activities and is calibrated with both synthetic datasets and actual mobility datasets provided by DiDi. Experiment results show that the proposed framework can significantly improve the system efficiency.
\end{itemize}

The rest of the paper is organized as follows. Section 2 describes the research problem, highlights the potential benefits of delayed matching, and points out the main challenges. Section 3 presents the proposed multi-agent reinforcement learning framework and algorithms. The simulator settings, experimental results and sensitivity analyses are demonstrated in Section 4, while the related work is presented in Section 5. Section 6 summarizes the study and outlook future research.

\section{Research problem}
\label{Sec: R_prob}

This section presents the relevant background knowledge and formulates a combinatorial optimization problem for the matching between unserved passengers and idle drivers. In a ride-sourcing system, when a passenger raises an on-demand ride request, he or she will be placed in a queue (or matching pool) and wait for being matched with a driver through the platform. The time from raising a request to being matched is referred to as the waiting time. After the passenger is matched with a driver, the driver should drive to the location of the passenger and pick him/her up (the time wasted in this process is denoted as pick-up time). When a passenger drops-on a vehicle and reaches the destination, or cancels his/her request halfway (during the waiting time or the pick-up time), the life cycle of his/her request is viewed as terminated. On the supply side, at each instant, each registered ride-sourcing driver has three status: off-line, occupied, and available (idle). Off-line status means that a driver closes his/her APP and is not willing to be dispatched by the platform, occupied status indicates that a driver has been already dispatched to a passenger (on the road to pick-up the passenger or to take the passenger to the destination), while the available status implies that a driver waits for being dispatched. 

With no doubt, the major concerns of the online matching are the unserved passengers, and available drivers. The goal of the online matching can be formulated as a bipartite matching problem between the unserved passengers and available drivers in the matching pool at each match time interval. To be specific, during each time interval, given $I$ unserved passengers $(i=1,2,\dots,I)$ and $J$ idle drivers $(j=1,2,\dots,J)$, the objective function and constraints for optimizing the online matching can be formulated as:

\begin{equation}\label{Eq1}
\begin{split}
    \min_{a_{ij}} & \sum_{i=1}^I \sum_{j=1}^J [c(i,j)-B] a_{ij} \\
    \text{s.t.} & \sum_{i=1}^I a_{ij} \leq 1, \ j = 1,2,\dots,J \\
    \ & \sum_{j=1}^J a_{ij} \leq 1, \ i = 1,2,\dots,I
\end{split}
\end{equation}
where $c(i,j)$ is the cost of assigning driver $j$ to passenger $i$, which is equal to the pick-up time from driver j to passenger i in this study. $a_{ij}$ are the index variables to be optimized, with the following definitions:
\begin{equation*}
    a_{ij} = 
    \begin{cases}
    1, & \text{if passenger}\ i\ \text{is matched with driver}\ j\\
    0, & \text{if passenger}\ i\ \text{is not matched with driver}\ j
    \end{cases}
\end{equation*}

And, $B$ is a large number which ensures at least one of the two constraints in Eq. \ref{Eq1} is binding. It implies that: if the number of available drivers is greater than the number of unserved passengers, then every unserved passenger will be matched (the first constraint is binding while the second constraint is non-binding); otherwise, then every available driver will be matched (the first constraint is non-binding while the second constraint is binding). 

\begin{figure}
  \centering
  \includegraphics[width=1\columnwidth]{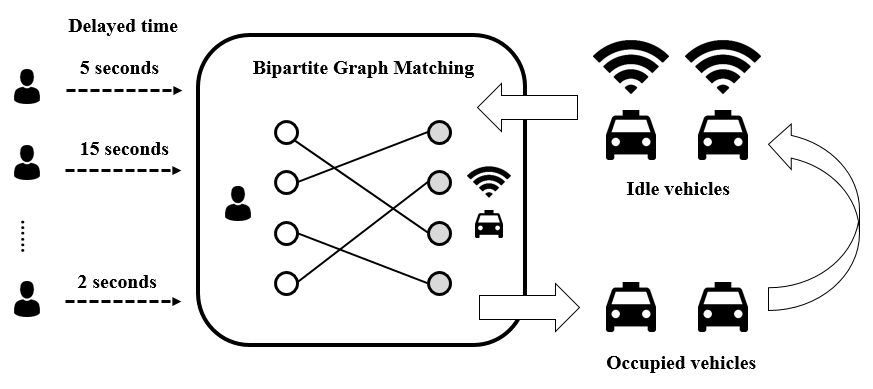}
  \caption{Online matching of a ride-sourcing system.}
  \label{Fig 2:Schematic}
\end{figure}

As aforementioned, a passenger request may experience a lower cost if its matching is delayed for a few match time intervals. Unlike most of the previous studies that focused on developing efficient bipartite graph matching algorithms or using reinforcement learning to control drivers' decisions, the main goal of this study is to propose a method that automatically determine the delayed time of each unserved passenger request, for better balancing the trade-offs among pick-up time, waiting time and matching success rate. Our proposed method is established on top of the bipartite graph matching and determines the time when each unserved request enters the matching pool to be matched with an idle driver. Fig. \ref{Fig 2:Schematic} illustrates the online matching procedure of a ride-sourcing system, which first determines the delayed time for each passenger and then implements the bipartite graph matching.

\section{A multi-agent learning framework}
\label{Sec: multi_agent}
This section first builds an MMDP in which each passenger request represents an agent, and the action of each agent during each match time interval is simplified as a binary choice - delayed or not delayed. Then the two reinforcement learning methods for optimal policy learning are presented. 

\subsection{Problem formulation}
The problem of determining each passenger’s delayed time is modelled as an MMDP Game, with the agents, states, actions, state transitions, and rewards defined as follows. 

\textbf{Agent}: we regard each passenger request as an agent, who emerges when it is raised and terminates when it is successfully matched with a driver or cancelled by the passenger. Clearly, the number of agents in each match time interval, denoted as $N_t$, may vary across different match time intervals $t$. This multi-agent based definition can greatly reduce the magnitude of the action space in comparison to the way that defines the centralized platform as an agent. 

\textbf{State}: states in each match time interval are spatio-temporal patterns of supply and demand that the agents observe. To be specific, the state for agent $i$ in match time interval $t$, $s_t^i$, consists of two components: the global-view state $s_{t,\left(G\right)}^i$, and the local-view state $s_{t,\left(L\right)}^i$. Suppose the whole examined area is partitioned into $M$ small grids (such as squares and hexagons), then global-view state vector $s_{t,\left(G\right)}^i$ includes four types of spatio-temporal supply-demand features: 1) the number of remaining unserved passengers left from the previous match time interval in each grid; 2) the number of remaining idle drivers left from the previous match time interval in each grid; 3) the expected arrival rate of new unserved passengers in each grid; 4) the expected arrival rate of new idle drivers in each grid;

The latter two types of features can be predicted with historical and real-time data. Clearly, global-view state vector is of $M\times4$ dimensions, i.e. $s_{t,\left(G\right)}^i\in \mathbf{R}^{M\times4}$. Evidently, the global-view state is shared by all agents in match time interval $t$. Note that more relevant spatio-temporal features can be incorporated into the global-view state vector.  The local-view state vector $s_{t,\left(L\right)}^i$ includes three components: 1) the location of agent $i$ that is represented with one-hot encoding; 2)  agent $i$’s cumulative waiting time; 3) the expected distance from agent $i$ to its matched driver ${\hat{T}}_t^i$, which is estimated according to a combinatorial optimization program shown in Eq.~\ref{Eq1}. Note that here the optimization program is run virtually for getting the features, i.e. ${\hat{T}}_t^i$ for all agents, rather than for actually matching unserved passengers and idle drivers.

Clearly, the local-view state is different across agents and has a dimensionality of $M+2$. Therefore, the state for agent $i$ during match time interval $t$ can be written as $s_t^i = \left[s_{t,\left(G\right)}, s_{t,\left(L\right)}\right] \in \mathbf{R}^{M\times5+2}$, then the joint state of all agents in match time interval $t$ can be represented $\boldsymbol{s}_t=\left\{s_t^1,s_t^2,\dots,s_t^{N_t}\right\}$.

\textbf{Action}: agent $i$ in match time interval $t$ has two available actions, $a_t^i \in \left\{0,1\right\}$, where $a_t^i=1$ denotes that the agent enters the matching pool where a bipartite matching is implemented, while $a_t^i=0$ indicates that the agent choose to not enter the matching pool and be delayed to the next match time interval. Then the joint action of all agents in match time interval $t$ can be written as  $\boldsymbol{a}_t=\left\{a_t^1,a_t^2,\ldots,a_t^{N_t}\right\}$. 

\textbf{State Transition}: When time moves from time interval $t$ to time interval $t+1$, the state of each agent gets updated according to the interactions between the agents’ actions and environment. First, the agents at time interval $t$ with action equal to 1 are assigned to idle drivers based on a combinatorial optimization algorithm. Clearly, some old agents are matched and leave the system. Second, the environment changes according to the simulator (which will be presented in the next section), and particularly, some new agents (passenger requests) may enter the system. Third, the states of the agents at time interval $t+1$ (including those unmatched agents and new agents) are calculated based on the current environments. It is worthy mentioned that the dimensionality of $\boldsymbol{s}_t$ is not necessarily equal to the dimensionality of $\boldsymbol{s}_(t+1)$, in other words, the number of agents in match time interval $t$ is not necessarily identical to that in match time interval $t+1$. This is because the number of matched agents at time interval $t$ is not necessarily equal to the number of newly coming agents at time interval $t+1$.

\textbf{Reward}: to examine the trade-offs on cooperation and competition among agents, two types of reward, i.e. global reward and individual reward, are considered. The individual reward of agent $i$ in match time interval $t$, denoted as ${\tilde{r}}_t^i$, is designed as follows. 
\begin{equation}\label{Eq2}
    {\tilde{r}}_t^i=
    \begin{cases}
    0, & \text{if}\ t<T_i\\
    V-c(i,j), & \text{if}\ t = T_i
    \end{cases}
\end{equation}

where $T_i$ is the last match time interval of agent $i$ (the life of an agent is terminated when it is matched or cancelled by the passenger). $V$ refers to the positive reward (benefit)  for successfully matching one pair of passenger and driver. Although this positive reward can depend on the characteristics of the request, such as the trip fare and expected trip time, we do not consider request discrimination and set a constant value $V$ for all successfully matched agents. To encourage agents to cooperate with each other, we further introduce a notion of global reward, which is defined as the average reward of all agents. Formally, the global reward assigned to agent $i$ in match time interval $t$, denoted as ${\bar{r}}_t^i$, equals to:

\begin{equation}\label{Eq3}
    {\bar{r}}_t^i=
    \begin{cases}
    0, & \text{if}\ t<T_i\\
   \frac{1}{N} \sum_i^N \left[ V-c(i,j)\right], & \text{if}\ t = T_i
    \end{cases}
\end{equation}

where $N$ refers to the number of agents that have appeared in the whole epoch. Notice that the average reward of all agents, i.e. $\frac{1}{N}\sum_i^N \left[ V-c(i,j)\right]$, is calculated at the end of the epoch (when the life of some agents has already terminated) and then assigned to the terminal match time interval $T_i$ for each agent. The final reward assigned to agent $i$ in match time interval $t$ can be written as:

\begin{equation}\label{Eq4}
    r_t^i = \rho \cdot \tilde{r}_t^i+(1-\rho)\cdot \bar{r}_t^i
\end{equation}

where $\rho \in [0,1]$ is a weighted factor that balance the tradeoff between the individual reward $\tilde{r}_t^i$ and global reward $\bar{r}_t^i$. Clearly, $\rho=1$ indicates that each agent aims to optimize its own reward instead of considering other agents’ reward, while $\rho=0$ implies that each agent tries to maximize the total reward of all agents. The goal of each agent $i$ is to maximize its own total expected discounted return $R^i=\sum_{t=0}^{T_i}\gamma^t r_t^i$ where $\gamma^t$ is a discount factor.

\subsection{Simulator}
One of the most important component for applying deep reinforcement learning methods is the learning environment, with which the agents interact. For training multi-agent reinforcement learning algorithms, many hand-crafted environments are designed \cite{lowe2017multi} \cite{mordatch2017emergence}, such as cooperative navigation where agents cooperate to reach various landmarks while avoiding colliding with each other, and predator-prey where some slower agents cooperate to chase the faster adversary. Multi-agent reinforcement learning methods are shown to achieve great performance in these human-designed games. However, implementation of these methods on real-world applications is much more difficult, due to the high dimensionality of action and state space, non-stationary environment, and improper reward design. The real-world environment is of high stochasticity and randomness, which makes the training and evaluation for the reinforcement learning algorithms difficult. Moreover, the training of reinforcement learning algorithms requires a large number of epochs, which may not be supported by limited historical data. One common solution is to build simulators which are calibrated with real historical data. 

In this section, we design a simulator that explicitly describes the dynamics in an online matching system. As shown in Algorithm \ref{Alg:simulator}, each match time interval includes the following steps: policy implementation of DRL that separates agents into two groups (entering the matching pool or being delayed to the next interval), bipartite matching between unserved passengers and idle drivers in the matching pool, update of drivers’ status due to completed trip and online/offline activities, and generation of new requests, etc.  In essence, the simulator iteratively updates the waiting list of unserved requests and the status of drivers (“idle”, “occupied” or “offline”). When a new request is generated, it is appended to the waiting list; when an unserved request is matched with an idle driver, it is removed from the waiting list. When a driver gets online (or offline), his/her status changes from “offline” to “idle” (or from “idle” to “offline”), respectively. When a driver is matched with a request or completes his/her trip, his/her status changes from “idle” to “occupied” (or from “occupied” to “idle”). Note that the number of agents $N^t$ may change with the match time interval since new agents join and some old agents may leave the waiting list of unserved requests. This setting is different from most of the previous studies related to multi-agent reinforcement learning, in which the number of agents is unchanged. It makes the design of the multi-agent learning algorithms more intractable. It is also worthy mentioned that the bipartite matching is executed virtually for estimating the expected pickup distance for agent $i$ while preparing the joint states for the next match time interval. 

\begin{algorithm}[htb]
    \caption{Simulator for an online matching system}\label{Alg:1}
    \label{Alg:simulator}
    \begin{algorithmic}[1]
        \parState {%
        \textbf{Input:} information of the historical passenger requests and distributions of drivers’ online/offline time.}
        \State \textbf{Initialize joint states}\ $\boldsymbol{s}_t$.
        \For {every match time interval of the online dispatching system ($t=0$\ to $T$)}
            \parState{%
            \textit{\textbf{Implement policies of DRL}}: Execute the joint actions $\boldsymbol{a}_t = \pi (\boldsymbol{s}_t)$ according to the policies of multi-agent DRL algorithms, where $\boldsymbol{a}_t$ determines whether each agent $i$ should enter the matching pool or be delayed to $t+1$.}
            \parState{%
            \textit{\textbf{Bipartite matching}}: Collect unserved passenger requests with action equal to 1 (i.e. entering the matching pool) and idle drivers, and execute bipartite matching according to Eq.~\ref{Eq1}.}
            \parState{%
            \textit{\textbf{Update matching outcomes}}: The matched requests are removed from the waiting list of unserved passenger requests; the status of the matched drivers are updated from “idle” to “occupied”.}
            \parState{%
            \textit{\textbf{Completed trips update}}: Update the status of drivers who completes their trips from “occupied” to “idle”.}
            \parState{%
            \textit{\textbf{Request generation}}: new passenger requests are bootstrapped from historical requests occurred in the same time period. The new requests are appended to the waiting list of unserved requests. Note that the number of agents $N_t$ changes after this step.}
            \parState{%
            \textit{\textbf{Online/offline status update}}: update the status of drivers who become online (available to serve) or offline (unavailable to serve) in the current interval, according to real historical distributions.}
            \parState{%
            \textit{\textbf{Generate next states}}: collect the spatio-temporal demand-supply conditions and update the state for each agent, outputting $\boldsymbol{s}_{t+1}$. Particularly, for estimating the expected distance from each agent i to its matched driver, we execute Eq.~\ref{Eq1} with the input of all unserved agents and idle drivers.}
    \EndFor
\end{algorithmic}
\end{algorithm}

\subsection{Reinforcement learning methods}

The objective of reinforcement learning is to identify a policy $\pi$ that achieves optimal accumulated reward for one agent or multiple agents. For single-agent applications, Q-learning is one widely used method. It estimates the expected total discounted rewards of state-action pairs, which can be approximated by a Q-function table that is solved with Bellman equation. A recent extension of Q-learning is Deep Q-network (DQN) which uses a neural network for approximating the Q-function table. DQN tries to minimize the difference between predicted Q-values and target Q-values that combines the current reward and estimated value of next state. The adaptation of Q-learning to multi-agent settings has been examined in the literature \cite{schwartz2014multi}, however, it is still an open question and the theoretical guarantees for multi-agent DQN remain unsolved. There are two main solutions: decentralized and centralized methods. In decentralized methods, each agent is characterized with separate neural network, and the only interaction among agents is the environment. For example,  \cite{tampuu2017multiagent}  established a decentralized DQN in Pong game environment where two (players) agents try to beat each other (in competition setting) or keep the ball bouncing between them (in cooperative setting). 

However, the decentralized methods are not well adapted to our problem for the two reasons below. First, the number of agents change with intervals, thus it is hard to identify the number of neural networks to be constructed ex ante. Second, there are numerous requests popping up and disappearing dynamically, indicating that the number of agents is extremely large. Modelling each agent with one neural network must require high amount of computational resources. Therefore, in this paper, we resort to centralized methods, which use one neural network to model the behaviors of all agents, in other words, the weights of the centralized neural network is shared by all agents. The parameter of the centralized network $\theta$ is updated by minimizing the total difference between the predicted value of the Q-networks and the target Q values, as shown in Eq.~\ref{Eq5}, in terms of the transitions of all agents $\left(s_t^i, a_t^i, r_t^i, s_{t+1}^i, d_t^i\right)$. As shown in Eq.~\ref{Eq6}, the target Q values are estimated by the obtained reward if the current state is the terminated state, and by the sum of the obtained reward and the discounted estimated value of next state otherwise.

\begin{equation}\label{Eq5}
    L(\theta) = E_{s_t^i, a_t^i, r_t^i, s_{t+1}^i}\Big[(Q(s_t^i, a_t^i\mid \theta)-y)^2\Big]
\end{equation}

\begin{equation}\label{Eq6}
    y = 
    \begin{cases}
    r_t^i, & \text{if}\ d_t^i = 1\\
    r_t^i+\gamma \max_{a_{t+1}^i} Q (s_{t+1}^i, a_{t+1}^i), & \text{if}\ d_t^i = 0
    \end{cases}
\end{equation}

The details of the algorithm of the proposed ST-M-DQN are shown in Algorithm~\ref{Alg:dqn}. To avoid instable training, the replay memory is used. In each match time interval, we first sample action for each agent with $\epsilon$-greedy policy based on the centralized Q network, then execute the joint actions in the simulator. The simulator will re-compute the spatio-temporal patterns of supply and demand and then returns rewards and next states for all agents, then the transitions of each agent $\left(s_t^i, a_t^i, r_t^i, s_{t+1}^i, d_t^i\right)$ are stored in the reply memory. For training the centralized Q-network, we randomly sample a mini-batch of transitions from the reply memory each time, and update the network parameter $\theta$ by minimizing the loss function in Eq.~\ref{Eq5}.

\begin{algorithm}
    \caption{ST-M-DQN}\label{Alg:dqn}
    \begin{algorithmic}[1]
        \State Initialize replay memory $D$; 
        \State Use random weights $\theta$ to initialize the action-value function.
        \For {$k=1$ to number of epochs}
            \parState {% 
            Reset the environment and obtain the initial joint states $\boldsymbol{s}_0$.}
            \For{every match time interval ($t=0$ to $T$)}
                \For{$i=1$ to $N_t$}
                    \parState {%
                    Sample action of each agent, $a_t^i$, according to $\epsilon$-greedy policy: $a_t^i=\argmax_{a_t^i} Q(s_t^i, a_t^i)$ with probability $1-\epsilon$ otherwise choose a random action.}
                \EndFor
                \parState {%
                Execute the simulator (Algorithm~\ref{Alg:1}) with the input of the joint actions $\boldsymbol{a}_t$, then observe joint reward $\boldsymbol{r}_t$ and next joint state $\boldsymbol{s}_{t+1}$. For agent $i$, we denote $d_t^i=1$ if its life is terminated, i.e. $t=T_i$, and $d_t^i=0$ otherwise.}
                \parState {%
                Store transitions of all agents $\left(s_t^i, a_t^i, r_t^i, s_{t+1}^i, d_t^i\right)$, $\forall i=1,\dots,N_t$ in the replay memory $D$.}
            \EndFor
            \For {$m=1$ to $M_1$}
                \parState {%
                Sample a mini-batch of transitions $\left(s_t^i, a_t^i, r_t^i, s_{t+1}^i, d_t^i\right)$, $\forall i=1,\dots,N_t$ from replay memory $D$.}
                \parState {% 
                Calculate target value according to Eq.~\ref{Eq6}.}
                \parState {%
                Update the parameters of Q-network by: $\theta \gets \theta+\beta_1\nabla_\theta L(\theta)^2$.}
            \EndFor
        \EndFor
\end{algorithmic}
\end{algorithm}

The second reinforcement learning method is the spatio-temporal multi-agent actor-critic (ST-M-A2C). Actor-critic (A2C) is a popular policy gradient method for reinforcement learning tasks. A2C establishes two networks, one policy network (known as critic and used to output policy) and one value network (known as actor and used to evaluate the performance of the policy network). It updates the parameters of the policy network $\theta_p$, and that of the value network $\theta_v$ iteratively. Similar to the DQN, due to the large and dynamically changing number of agents, we design a centralized multi-agent A2C. The weights of the centralized value network $\theta_v$ are shared across all agents, and the update of $\theta_v$ can be achieved by minimizing a loss function presented in Eq.~\ref{Eq7}, where $V_{\theta_v}(S_t^i)$ is the predicted value of value network and $V_{\text{target}}(S_{t+1}^i;\theta_v^\prime,\pi)$ is the target value constituted of immediate reward and discounted estimated value of next state, as shown in Eq.~\ref{Eq8}. 

\begin{equation}\label{Eq7}
    L(\theta_v) = \Big[V_{\theta_v}(S_t^i)-V_{\text{target}}(S_{t+1}^i;\theta_v^\prime,\pi)\Big]
\end{equation}

\begin{equation}\label{Eq8}
    V_{\text{target}}(s_{t+1}^i;\theta_v^\prime,\pi) = \sum_{a_t^i} \pi (a_t^i \mid s_t^i) \Big[r_t^i+ \gamma V_{\theta_v^\prime(s_{t+1}^i)}\Big]
\end{equation}

With the parameters of the value network $\theta_v$, ST-M-A2C updates the parameters of the policy network $\theta_p$ by a gradient descent rule $\theta_p \gets \theta_p + \beta_2 \nabla_{\theta_p} J(\theta_p)$, where $\beta_2$ is the learning rate for actor, and the gradient $\nabla_{\theta_p} J(\theta_p)$ is given by Eq.~\ref{Eq9}. To reduce high variability of value functions, an advantage function estimated by the TD-error defined in Eq.~\ref{Eq10} rather is used for calculating the policy gradient. 
\begin{equation}\label{Eq9}
    \nabla_{\theta_p} J(\theta_p)  = \nabla_{\theta_p} \log \pi_{\theta_p} (a_t^i \mid a_t^i) A (s_t^i, a_t^i)
\end{equation}

\begin{equation}\label{Eq10}
    A(s_t^i, a_t^i) = a_{t+1}^i+\gamma V_{\theta_v^\prime}(s_{t+1}^i) - V_{\theta_v}(s_{t}^i)
\end{equation}

The details of the ST-M-A2C is illustrated in Algorithm \ref{Alg:a2c}.

\begin{algorithm}
    \caption{ST-M-A2C}\label{Alg:a2c}
    \begin{algorithmic}[1]
        \State Initialize replay memory $D$; 
        \State Use random weights $\theta_v$ to initialize the value network.
        \For {$k=1$ to number of epochs}
            \parState {% 
            Reset the environment and obtain the initial joint states $\boldsymbol{s}_0$.}
            \For{every match time interval ($t=0$ to $T$)}
                \For{$i=1$ to $N_t$}
                    \parState {% 
                    Sample action of each agent, $a_t^i$ based on the action probability $P(s_t^i)$.}
                \EndFor
                \parState {%
                Execute the simulator (Algorithm~\ref{Alg:1}) with the input of the joint actions $\boldsymbol{a}_t$, then observe joint reward $\boldsymbol{r}_t$ and next joint state $\boldsymbol{s}_{t+1}$.}
                \parState {% 
                Store transitions of all agents $\left(s_t^i, a_t^i, s_{t+1}^i\right)$, $\forall i=1,\dots,N_t$ in the replay memory $D$.}
            \EndFor
            \For {$m=1$ to $M_2$}
                \parState {% 
                Sample a mini-batch of transitions $\left(s_t^i, a_t^i, s_{t+1}^i\right)$, from the replay memory $D$.}
                \parState {% 
                Update the parameters of value network $\theta_v$ by maximizing Eq.~\ref{Eq7}.}
                \parState {%
                Compute the advantage $A(s_t^i, a_t^i)$ by Eq.~\ref{Eq10} and update the parameters of policy network $\theta_p$ by $\theta_p \gets \theta_p + \beta_2\nabla_{\theta_p} L(\theta_p)$, where $\nabla_{\theta_p} L(\theta_p)$ is calculated with Eq.~\ref{Eq9}.}
            \EndFor
        \EndFor
\end{algorithmic}
\end{algorithm}

\section{Experiments}
\label{Sec: exp}
In this section, extensive experiments and sensitivity analyses are conducted to evaluate the performance of the proposed methods and investigate the impacts of the key parameters. 

\subsection{Experiments on a customized environment}
Here we first design a customized environment for illustrating the effectiveness of the proposed methods. We consider an $4\ \text{km}\times4\ \text{km}$ area with 30 match time intervals (each interval equals $1\ \text{s}$), and a simplified scenario the waiting passengers arrival at a rate of $q^d$ (in unit/s), while the idle drivers arrival at a rate of $q^s$ (in unit/s). Meanwhile, new arrival waiting passengers’ and idle drivers’ are generated based on a two-component mixture of Gaussians in each match time interval. The mean and standard deviation of arrival locations of waiting passengers are ($1.2 \ \text{km}$, $1.2 \ \text{km}$) and ($0.8 \ \text{km}$, $0.8 \ \text{km}$), while the mean and standard deviation of arrival locations of idle drivers are ($2.8 \ \text{km}$, $2.8 \ \text{km}$) and ($0.8 \ \text{km}$, $0.8 \ \text{km}$). Without considering the traffic congestions, we assume a constant vehicle speed as $25 \ \text{km/h}$. The road distance between any two points is estimated by the Manhattan distance, and given the constant vehicle speed, then the pickup time between each pair of waiting passenger and idle driver can be identified. The whole examined area is uniformly partitioned into $10\times10$ zones. Then the spatio-temporal features or states can be extracted by counting the number of idle drivers and waiting passengers in each zone during each time interval. Note that the partitioned zones are only used for perceiving the spatio-temporal features. Other features, such as the agent’s expected pickup distance, are calculated based on a continuous space.  

The two proposed approaches, ST-M-DQN and ST-M-A2C, are compared with two benchmarks: pure optimization strategy and tabular Q-learning. Pure optimization strategy matches idle drivers and waiting passengers immediately without considering any delayed matching with a combinatorial optimization depicted in Eq.~\ref{Eq1}. In this simplified customized environment, we do not consider the request cancelling behaviors of passengers, and thus the answer rate (proportion of the successfully matched passengers) is always equal to 1 if supply (idle drivers) is larger than or equal to demand (waiting passengers). The tabular Q-learning learns a Q-table that maps the states to action, with $\epsilon$-greedy policy. In tabular Q-learning, the state only includes the location and time of the agents, and thus the Q-table has a dimension of $T\times N\times 2$, where $N$ is the total number of separated zones.

Three metrics are used for comparing the effectiveness of the proposed models and benchmarks: mean reward of each agent ($R^i$ for agent $i$), answer rate and mean pickup time. The reward function is specified as follows. Let the positive reward (benefit) for successfully matching one passenger-driver pair, $V=800s$. It is noteworthy that $V$ is a decision variable determined by the platform and reflects the trade-offs between different objectives. A large $V$ indicates that the platform focuses more on successfully matching drivers and passengers with less considerations on the cost of pickup time, and vice versa. Therefore, the decision variable $V$ can be adjusted according to the objectives of different platforms. To compare the effectiveness of the models under different environments, three environmental settings are implemented: $q^d$=$q^s$=1 unit/s, $q^d$=$q^s$=2 unit/s, and $q^d$=$q^s$=3 unit/s. Both the value function approximation networks and policy networks are three-layer networks, with 512, 256, and 128 neurons from the first to last hidden layer. The activations of all hidden units are ReLu, while output layers of the value function approximation networks and policy networks use Linear and Softmax activations, respectively. All the experiments are repeated by 5 times to ensure the robustness of the results.

\begin{figure*} 
    \centering
  \subfloat[Mean reward of agent\label{3a}]{%
       \includegraphics[width=0.33\linewidth]{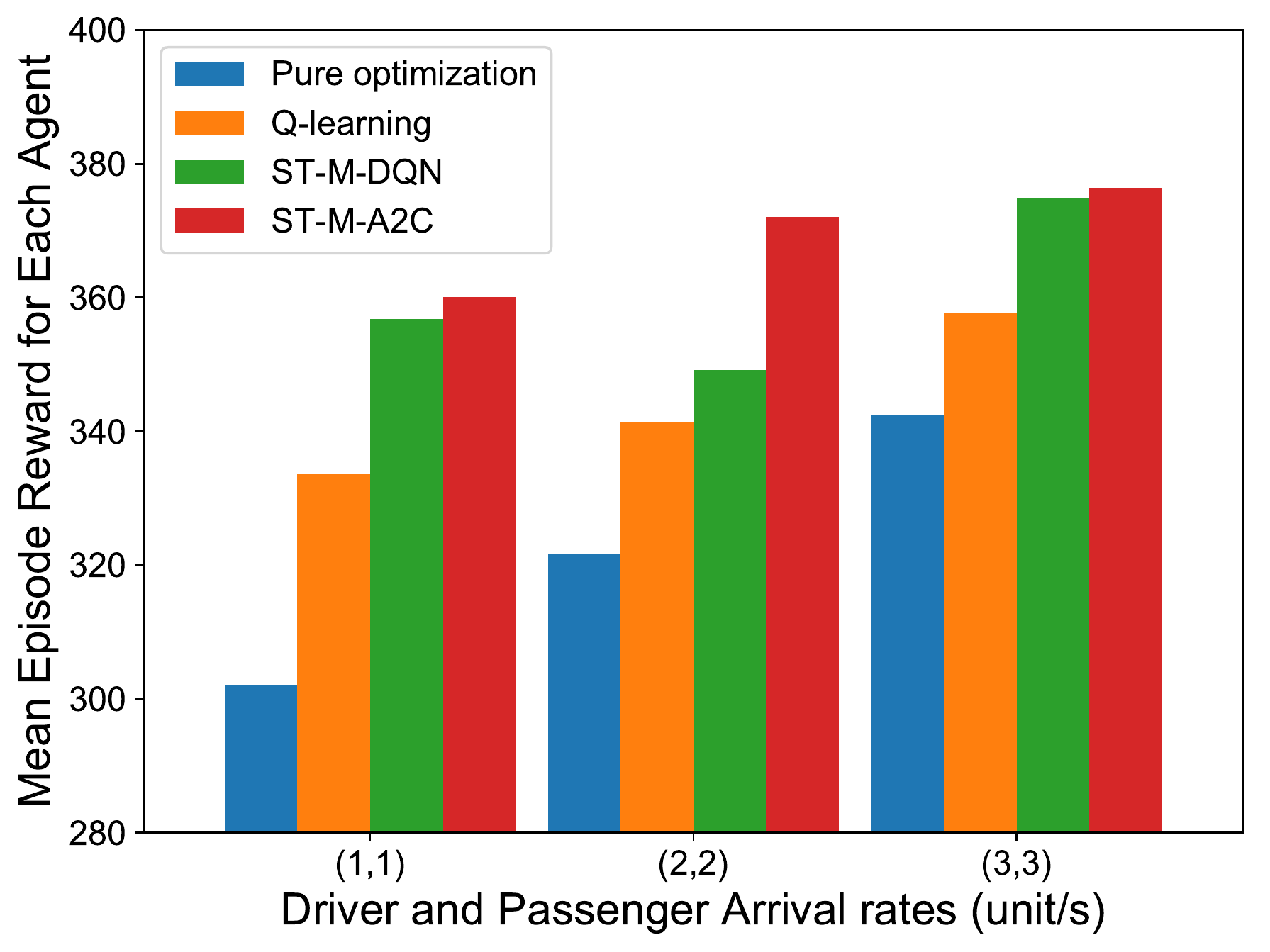}}
       \hfill
  \subfloat[Mean pickup time\label{3b}]{%
        \includegraphics[width=0.33\linewidth]{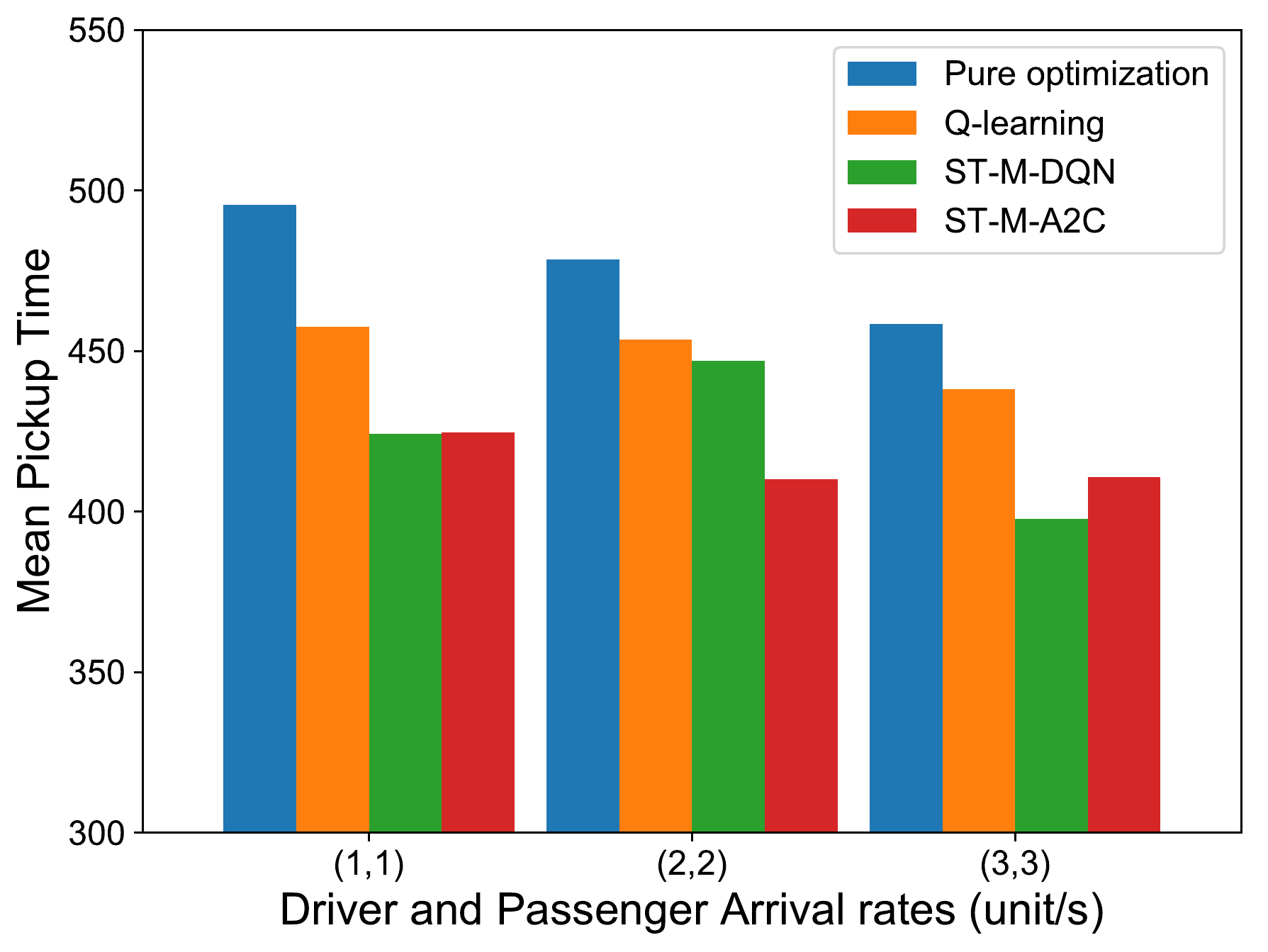}}
    \hfill
  \subfloat[Answer rate \label{3c}]{%
        \includegraphics[width=0.33\linewidth]{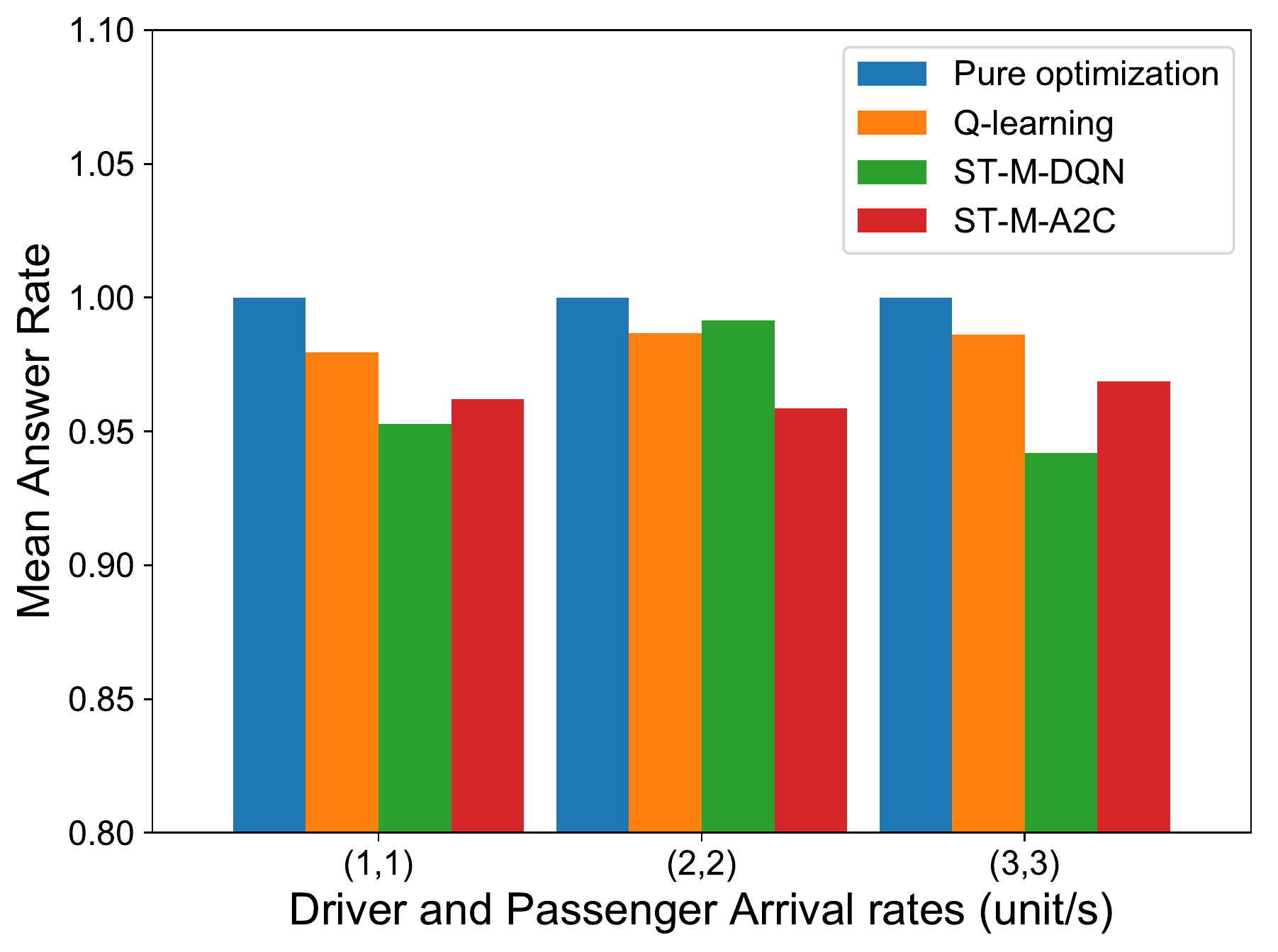}}
  \caption{Model comparisons on a customized environment.}
  \label{Fig 3:model} 
\end{figure*}

Fig. \ref{Fig 3:model} shows the resulting mean reward of each agent, answer rate, and mean pickup time of Pure optimization, Q-learning, ST-M-DQN, ST-M-A2C on the three environments. The results show that the mean reward of each agent increases with the arrival rate of drivers and passengers. This is intuitive since a better bipartite matching (with lower mean pickup time) becomes available as the increase of the density (dominated by the arrival rates) of drivers and passengers. In addition, it can be observed that the two proposed reinforcement learning methods significantly outperform the pure optimization and the tabular Q-learning methods in terms of the mean reward of each agent (which is the objective of the algorithms). The ST-M-A2C achieves the best performance in all environmental settings, which demonstrates its robustness. From Fig. \ref{Fig 3:model} (b)-(c), we find that the delayed matching controlling by the two deep multi-agent reinforcement learning methods can significantly reduce the mean pickup time of passenger-driver pairs with little loss on the answer rate. For example, when $q^d$=$q^s$=1 unit/s, the implementation of ST-M-A2C reduced the mean pickup time by 14.3\% while decreasing the mean answer rate by 3.8\%, resulting in 19.1\% of increase in agents’ mean reward. 

\begin{figure*} 
    \centering
  \subfloat[Arrival rates $q^d=q^s=1$ unit/s \label{4a}]{%
       \includegraphics[width=0.33\linewidth]{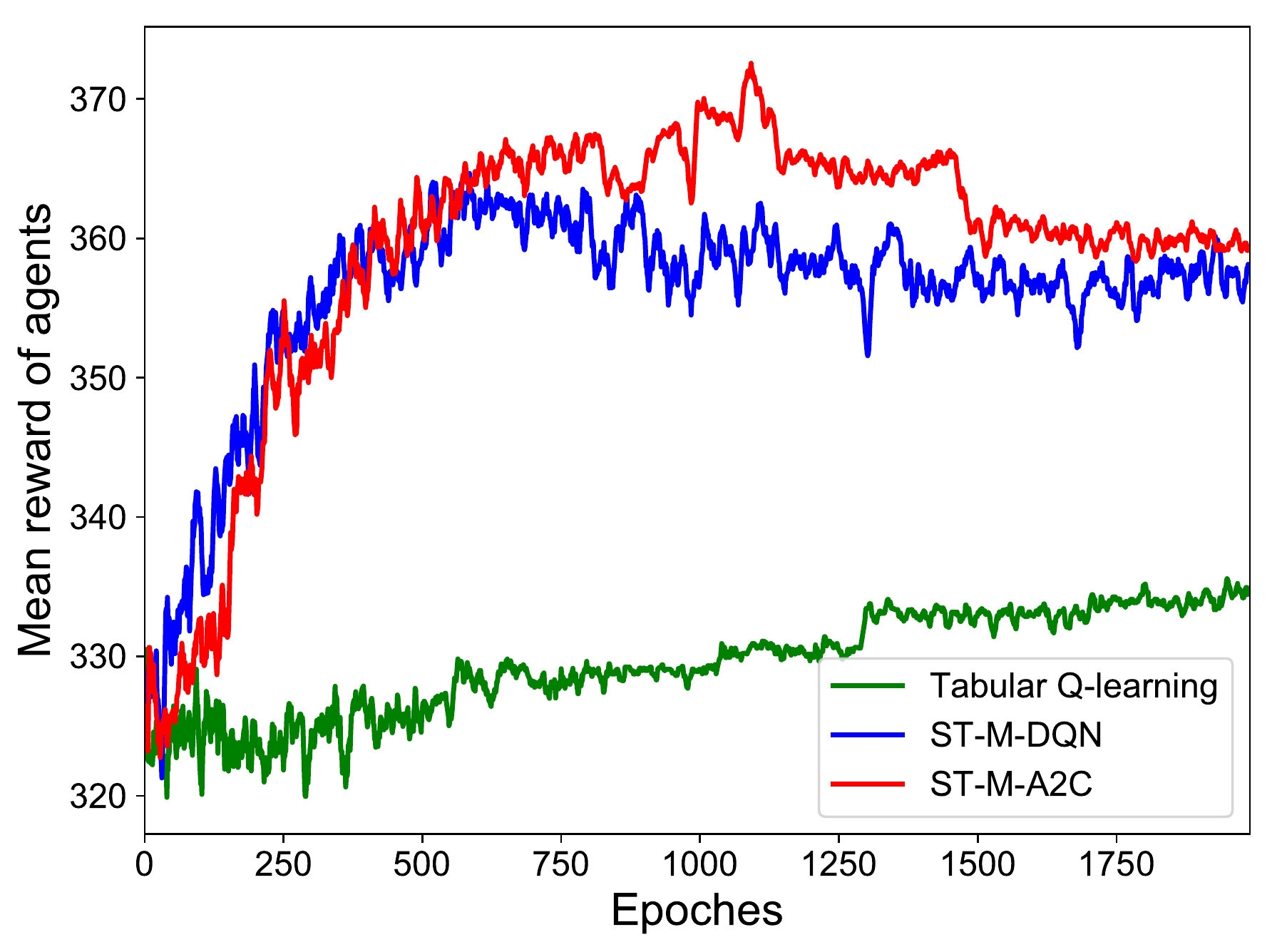}}
       \hfill
  \subfloat[Arrival rates $q^d=q^s=2$ unit/s \label{4b}]{%
        \includegraphics[width=0.33\linewidth]{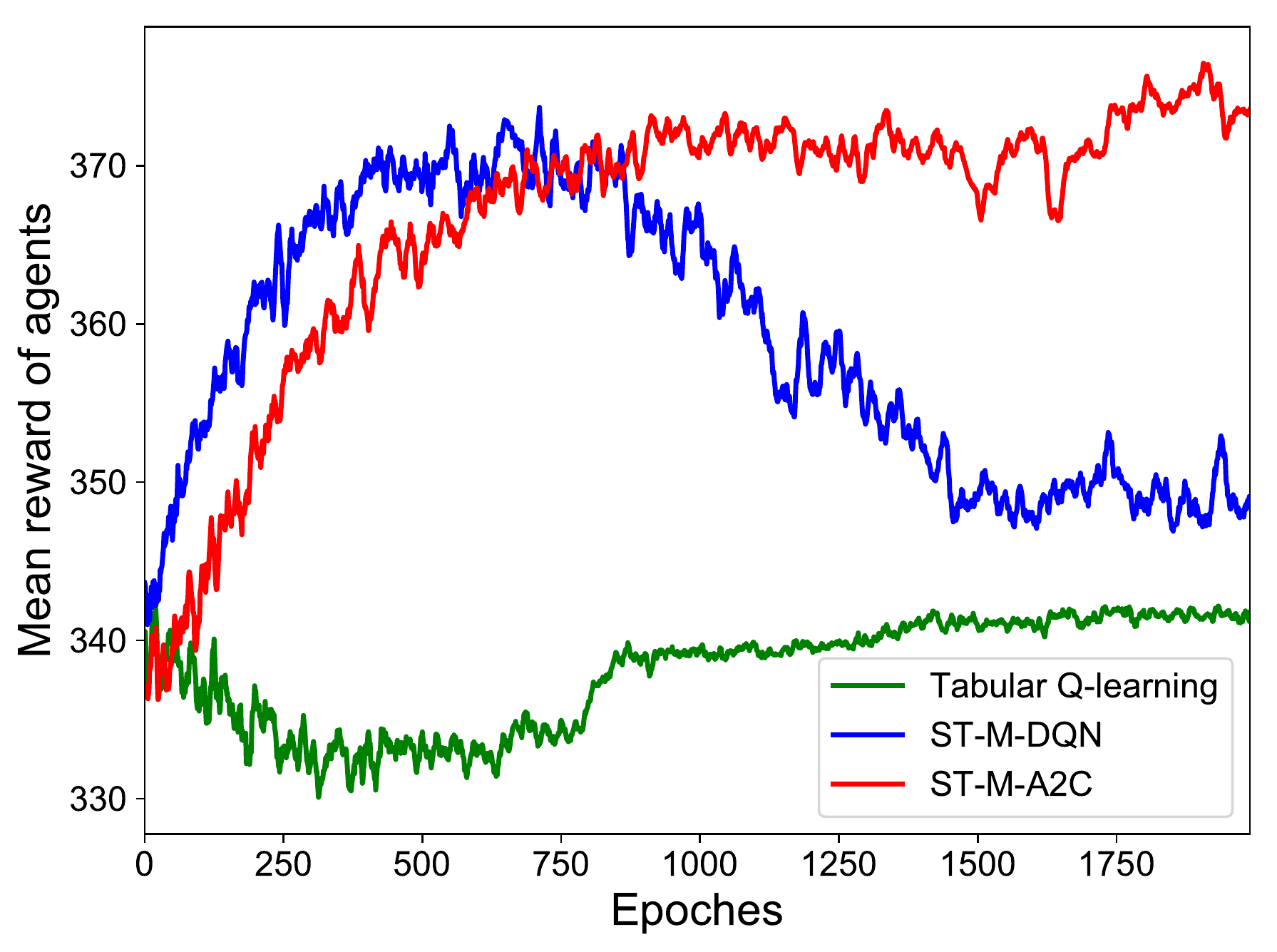}}
    \hfill
  \subfloat[Arrival rates $q^d=q^s=3$ unit/s \label{4c}]{%
        \includegraphics[width=0.33\linewidth]{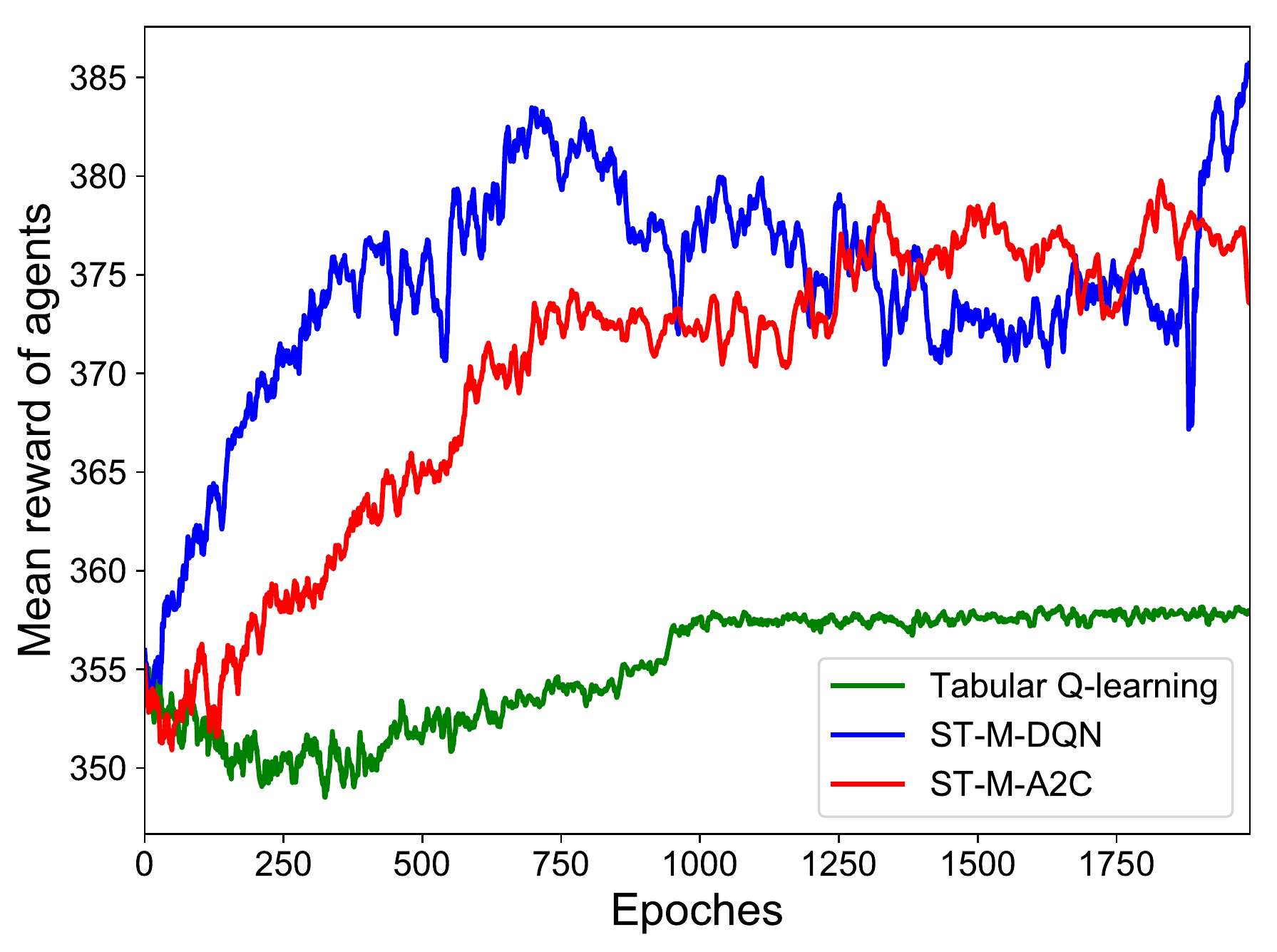}}
  \caption{Convergence curves of different models.}
  \label{Fig 4:convergence} 
\end{figure*}

Fig. \ref{Fig 4:convergence} shows the trends of average reward of agents with respect to training epochs of the three examined reinforcement learning models under different environment settings. Results show that ST-M-A2C and ST-M-DQN significantly outperform Tabular Q-learning in terms of converging speed, converged agents’ reward, and robustness. In general, ST-M-A2C shows more robust training curves and achieves higher converged agents’ reward than ST-M-DQN. 

\subsection{Sensitivity analysis in terms of reward weighted factor}

Next we conduct sensitivity analysis to look at the effects of reward weighted factor $\rho$ for the training of the proposed deep multi-agent reinforcement learning methods. As aforementioned, $\rho$ measures the tradeoff between individual reward $\tilde{r}_t^i$  and total reward $\bar{r}_t^i$i. The higher the $\rho$, the more emphasis that each agent pays to its individual reward versus total reward of all agents. It is naturally expected that using the total reward as signals for agents has potential to achieve better overall system performance. However, as \cite{sunehag2017value} mentioned, the use of total reward generates weak incentives to each agent, and may make some agents “lazy” to obtain rewards. Consequently, the resulting total rewards of all agents can be even worse than the algorithms using individual reward for each agent. Here, we train and test ST-M-A2C under the environmental settings of $q^d=q^s=1\ \text{unit/s}$, given five groups of reward weighted factor $\rho$ (from 0.0 to 1.0 with a step of 0.25).

\begin{figure}
  \centering
  \includegraphics[width=1\columnwidth]{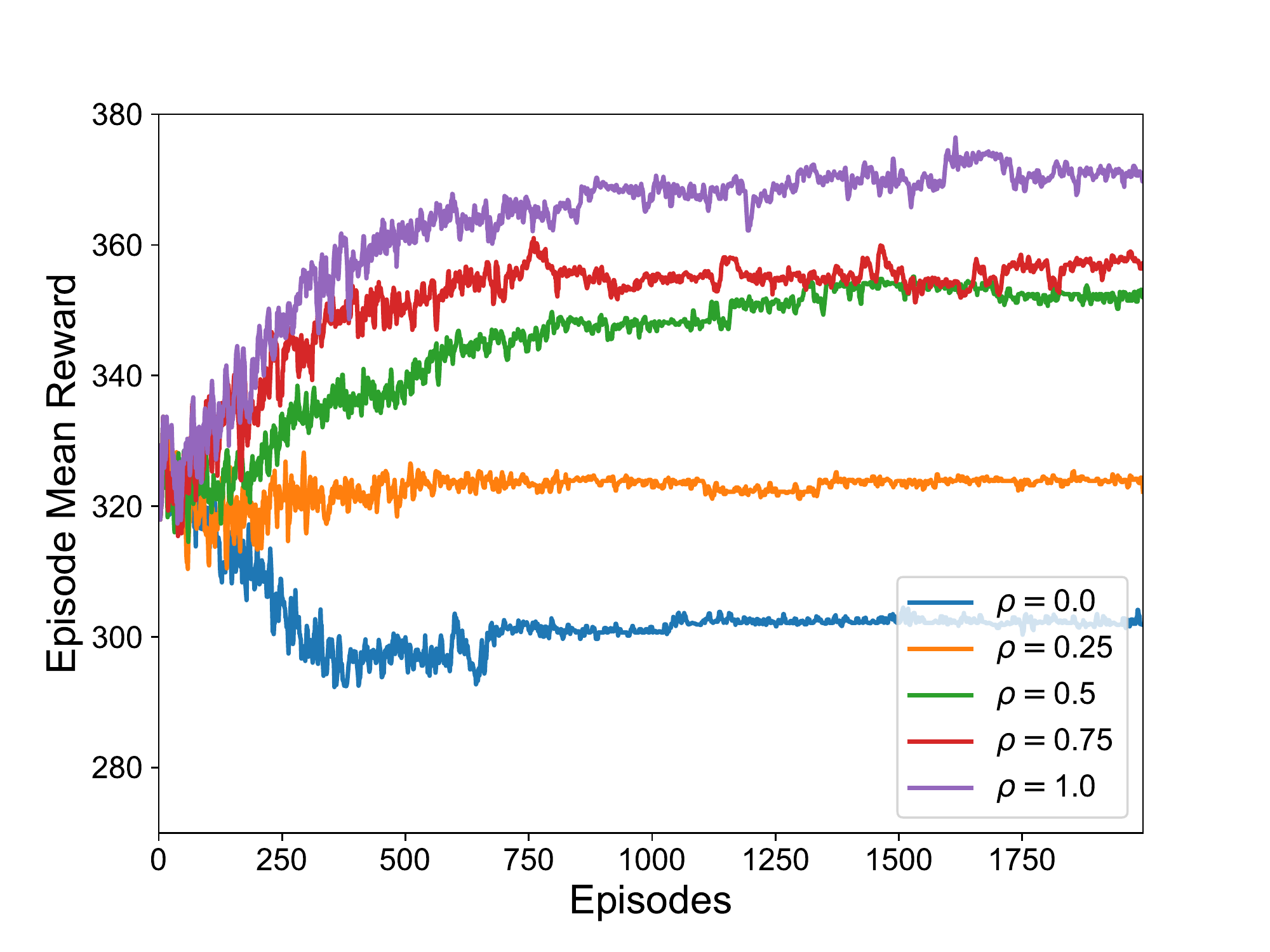}
  \caption{Evolution of episode mean reward under different reward weighted factor $\rho$}
  \label{Fig 5:evolution}
\end{figure}

Table \ref{Table:customized} demonstrates the answer rate, average pickup time and average agents’ reward of ST-M-A2C given different reward weighted factor $\rho$. Fig. \ref{Fig 5:evolution} further plots the training trends of the agents’ average reward with episodes under different reward weighted factors. The results show that the effectiveness and robustness of ST-M-A2C increases with the reward weighted factor $\rho$, indicating that the multi-agent framework behaves better when they are rewarded by their individual rewards in our problem. The possible reason is that the global reward (the sum of rewards of all agents) may be not strong enough to guide each agent and may encourage the “lazy” behaviors of agents. Those “lazy” agents do not learn from trial and error but simply select action 1 (not-delayed action) when they interact with any environment states. This observation is also consistent to the previous studies in the domain of taxi order dispatching or fleet management \cite{xu2018large} \cite{lin2018efficient} which normally use individual rewards as incentives for each agent in the multi-agent environments. 

\begin{table}[!t]
	\caption{{Model performances under different reward weighted factors}}
	\label{Table:customized}
	\centering
	\scriptsize
	\begin{tabularx}{0.5\textwidth}{lcccc}
		\toprule
		\toprule
		Model & \shortstack{Weighted  \\ factor $\rho$} & \shortstack{Answer \\ rate} & \shortstack{Mean pickup \\ time (s)} & \shortstack{Average \\ reward (s)} \\
		\midrule
		ST-M-A2C & 0 & 0.997 & 494.68 & 302.36 \\
		ST-M-A2C & 0.25 & 1.000 & 474.34 & 323.88 \\
		ST-M-A2C & 0.5 & 0.986 & 439.11 & 352.70 \\
		ST-M-A2C & 0.75 & 0.986 & 436.37 & 355.70 \\
		ST-M-A2C & 1 & 0.944 & 405.65 & 371.13 \\
		Pure Optimization & - & 1.000 & 495.56 & 302.17 \\
		\bottomrule
		\bottomrule
	\end{tabularx}
\end{table}

\subsection{Model performances on a real environment}

Apart from the customized environment, we further evaluate the performances of the proposed methods based on a simulator calibrated with real mobility data. The dataset, provided by the largest ride-sourcing platform in China, Didi Chuxing, includes four-week orders and drivers’ information in the downtown area ($20 km \times 20 km$) of a city of China. As shown in Algorithm \ref{Alg:simulator}, the input of the simulator includes a table of passenger requests and a table of drivers’ status. The former table contains the following trip-based attributes: the time and location (longitude and latitude) at which the passenger requests her order, the destination location, and the trip duration (from the time at which the passenger gets aboard to the time at which she drops off), and passenger ID. These attributes are extracted from the real-happened trips. The later table records the following driver-based attributes: the time and location at which the driver gets online and starts listening to orders, the time at which she gets offline and terminates her work, and driver ID. These attributes of drivers are also retrieved from the actual ride-sourcing drivers who register with the platform. With these two calibrated tables as input, the simulator with Algorithm \ref{Alg:simulator} can be executed. The performances in terms of average pickup time, answer rate and reward are evaluated on the four models, including the two proposed models, ST-M-DQN, ST-M-A2C, and two baselines, Pure optimization, Q-learning. In terms of spatio-temporal features, we partition the examined area into $20 \times 20$ zones, each of which occupies an area of 1 square kilometer. At each step, the number of idle drivers and the number of waiting passengers within each zone are counted and fed into the deep reinforcement models as spatio-temporal features. 

\begin{figure}
  \centering
  \includegraphics[width=1\columnwidth]{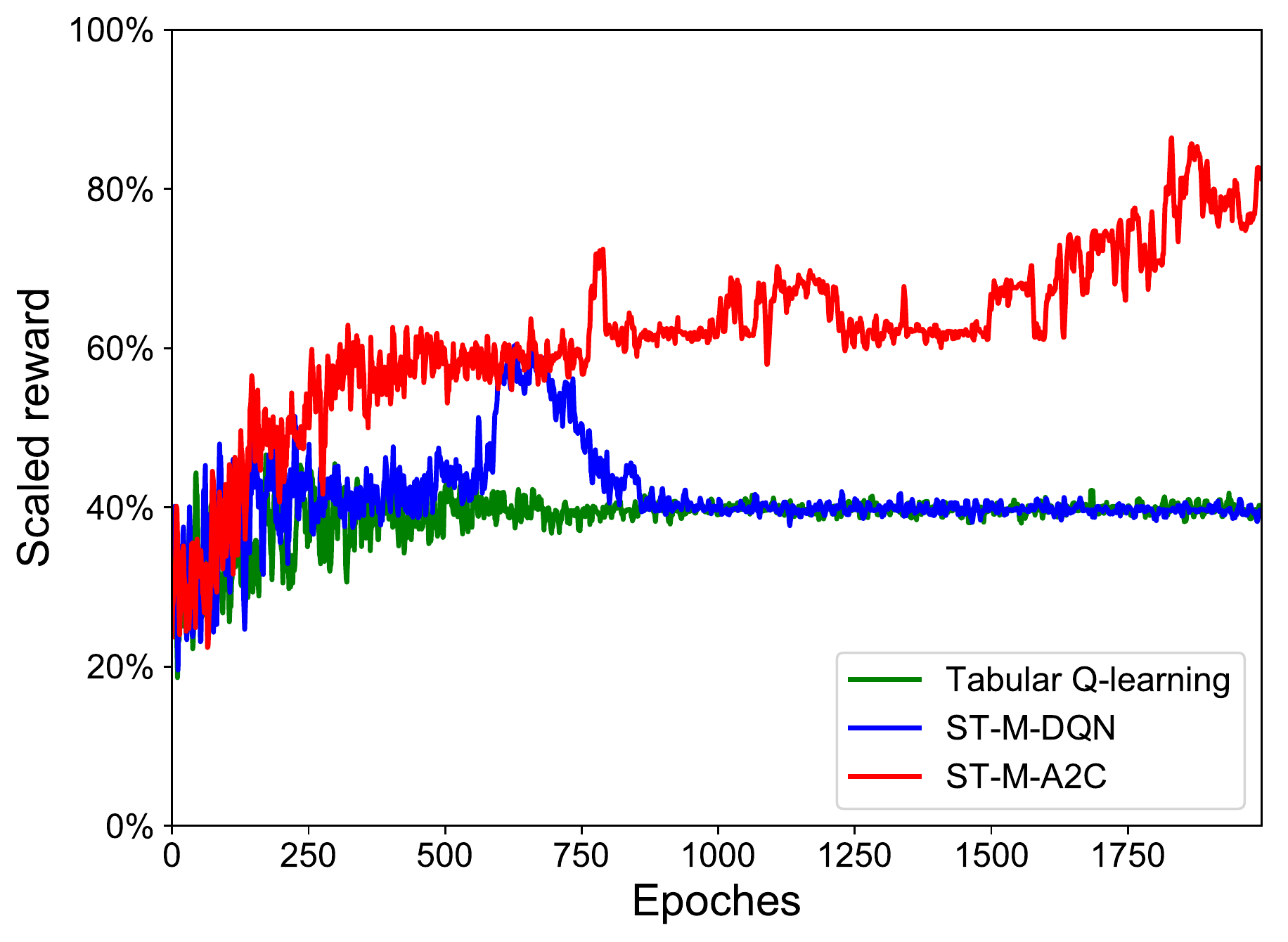}
  \caption{Evolution of episode scaled reward in real environment}
  \label{Fig 6:evolution}
\end{figure}

Table \ref{Table:real} shows the performance measures (answer rate, mean pickup time and mean reward of agents) of the three reinforcement learning models compared to pure optimization. Fig. \ref{Fig 6:evolution} shows the evolution of the mean reward of agents (which is scaled to a range from 0 to 1 due to privacy concerns) with episodes. It can be observed ST-M-A2C has the best performance; it significantly reduces the mean pickup time by 12.93\% with minor decrease on answer rate (by 0.46\%), and as a result achieves 6.02\% higher mean reward of agents than pure optimization. However, neither of ST-M-DQN and Tabular Q-learning has a good performance in an environment calibrated by real historical data. From Fig. \ref{Fig 6:evolution}, we can also observe that the training of ST-M-A2C is more robust and effective than ST-M-DQN and Tabular Q-learning. As a supplement to the experiments on a customized environment, this experiment on a simulator calibrated by real mobility data further evaluates the effectiveness of the proposed methods.

\begin{table}[!t]
	\caption{{Effectiveness of models in comparison to pure optimization}}
	\label{Table:real}
	\centering
	\scriptsize
	\begin{tabularx}{0.5\textwidth}{lcccc}
		\toprule
		\toprule
		Agent & \shortstack{Increases in   \\ answer rate} & \shortstack{Increases in \\ mean pickup time} & \shortstack{Increases in \\ average reward} \\
		\midrule
		Tabular Q-learning & -0.78\% & -1.408\% & 0.098\% \\
		ST-M-DQN & -0.508\% & -1.318\% & 0.108\% \\
		ST-M-A2C & -0.468\% & -12.938\% & 6.028\% \\ 
		\bottomrule
		\bottomrule
	\end{tabularx}
\end{table}

\section{Related work}
\label{Sec: related_work}
\textbf{Taxi dispatch:} Taxi dispatch, or driver dispatch, is a term usually referring to the process of matching vacant drivers with passengers’ requests using some algorithms to maximize the system’s performance. Traditional dispatch systems maximize the driver acceptance rate for each individual order by sequentially dispatching taxis to riders. \cite{zhang2017taxi} proposed to dispatch taxis to server multiple bookings at the same time thus to maximize the global success rate. \cite{wang2017stable} considered the individual participant’s benefit and proposed a notion of a stable match. \cite{wang2017deepsd} constructed an end-to-end framework to predict the future supply and demand in order to optimally schedule the drivers in advance. \cite{zhang2017taxi} investigated the preferred service and proposed a recommendation system to enhance the prediction accuracy and reduce the user’s effort in finding the desired service.

\cite{xu2018large} proposed an order dispatch algorithm that combined an offline learning step and an online planning step. The offline learning step estimated the value functions for the spatio-temporal patterns of passenger demand and taxi supply, then the online planning step executed an optimal matching between drivers and orders with the learned value functions. Their main focus was to enhance the traditional combinatorial optimization algorithm, such as Kuhn-Munkres (KM) algorithm, by adding the action-space values to the objective function. \cite{lin2018efficient} proposed a multi-agent reinforcement learning framework to tackle the fleet management problem that relocated the idle taxis to improve taxi utilization rate. \cite{li2019efficient} proposed a Mean Field Multi-Agent Reinforcement Learning for order dispatching in Didi's platform. They found that the mean field approximation was able to globally capture the demand and supply dynamics by propagating many local interactions between agents and the environment. \cite{wang2018deep} combined a deep Q-network with transfer learning techniques in a large-scale online order dispatching system. They showed that the strategies learned by knowledge transfer from sources city to target cities remarkably improved the system efficiencies, compared to strategies without transfer learning. However, as aforementioned, none of these studies have investigated the impacts of matching time intervals as well as the potential benefits of delayed matching. 

\textbf{Deep Reinforcement Learning:} Deep Reinforcement Learning (DRL) is a rapidly developed field which has attracted great attention especially since the emergence of AlfaGo \cite{silver2016mastering,silver2017mastering}. DRL combines the advantages of both deep learning (DL) and reinforcement learning (RL), where DL learns abstract or hidden features from large-scale data with the capability of using multiple processing layers \cite{lecun2015deep} and RL enables agent learning by interacting with its environment. Regarding the application of DRL in the transportation system, it is a multi-agent problem in the high-dimensional and non-stationary space. The powerful DRL provides a possible way to solve the multi-agent system problem, which is conventionally encountered in a variety of domains including robotics, control and communication \cite{busoniu2008comprehensive}. \cite{tampuu2017multiagent} evaluated the cooperation and competition between different agents when they share same environment. \cite{ye2015multi} investigated the wireless sensor networks using a multi-agent framework and found that cooperative neighbors effectively help the sensor to relay packets.

\section{Conclusions}
\label{Sec: conclu}

This study proposes a two-stage framework for online matching in ride-sourcing systems. The lower part contains a traditional convex combinatorial optimization algorithm that matches idle drivers and waiting passengers in the matching pool with minimum cost measured by pickup time. The upper part establishes two multi-agent deep reinforcement learning models, named ST-M-DQN and ST-M-A2C, which serve as multiple gates to control whether an agent should enter the matching pool in each matching time interval. These reinforcement learning models essentially determine the delayed time of each agent and help improve the effectiveness of the matching process, while the potential benefits of delayed matching are discussed in the literature. Through extensive experiments, we show that the proposed ST-M-DQN and ST-M-A2C well balance the trade-off between pickup time and matching rate and significantly improve the matching effectiveness in comparison to pure optimization and other benchmarks. It shows that the delayed matching controlled by the reinforcement learning methods can indeed remarkably reduces the average pickup time by incurring little loss on answer rate. This paper provides a novel framework that combines reinforcement learning and bipartite matching and evaluate its effectiveness in a ride-sourcing online matching system.

% \appendices
% \section{Notations}

\ifCLASSOPTIONcompsoc
  \section*{Acknowledgments}
\else
  \section*{Acknowledgment}
\fi

The work described in this paper was supported by Hong Kong Research Grants Council under projects HKUST16222916, NHKUST627/18 and the National Natural Science Foundation of China under projects 71622007, 7181101024.

\ifCLASSOPTIONcaptionsoff
  \newpage
\fi

\end{document}